\documentclass[fleqn,twoside]{article}
\usepackage{espcrc2}
\usepackage{graphicx}
\usepackage{overpic}
\usepackage{epsfig}
\usepackage{pstricks}
\usepackage[figuresright]{rotating}

\def\defineTColor#1#2{
  \newpsstyle{#1}{%
    fillstyle=vlines,hatchcolor=#2,%
    hatchwidth=0.1\pslinewidth,hatchsep=1\pslinewidth%
}}
\defineTColor{tCyan}{cyan}
\newrgbcolor{cyan}{0.878 1. 1.}
\newcommand{\myframebox}[3]{%
\psframebox[fillstyle=solid,fillcolor=#2]{%
\parbox{#1\textwidth}{%
       \black \begin{flushleft}{#3}\end{flushleft}
        }}}

\newcommand{\CK}{\v Cerenkov}

\newcommand{\AmS}{{\protect\the\textfont2
  A\kern-.1667em\lower.5ex\hbox{M}\kern-.125emS}}

\title{The Ring Imaging Cherenkov detector of the AMS experiment: test beam
  results with a prototype}

\author{Lu\'{i}sa Arruda\address[LIP]{LIP/IST \\
         Av. Elias Garcia, 14, 1$^{\textrm{\footnotesize{o}}}$ andar\\
         1000-149 Lisboa, Portugal \\
         e-mail: luisa@lip.pt},
  \thanks{I wish to thank the Funda\c {c}\~{a}o para a Ci\^{e}ncia e a Tecnologia
         for all the financial support for the journey to Siena and I would
         like to thank the conference organizers for the grant received to
         attend the meeting, as well as, for the excellent organization.}
 Fernando Bar\~ao\addressmark, Patr\'icia Gon\c calves\addressmark, Rui Pereira\addressmark}
       
\begin{document}

\begin{abstract}
The Alpha Magnetic Spectrometer (AMS) to be installed on the International
Space Station (ISS) will be equipped with a proximity Ring Imaging Cherenkov
(RICH) detector for measuring the velocity and electric charge of the charged
cosmic particles. This detector will contribute to the high level of
redundancy required for AMS as well as to the rejection of albedo
particles. Charge separation up to iron and a velocity resolution of the
order of 0.1\% for singly charged particles are expected. A RICH protoptype
consisting of a detection matrix with 96 photomultiplier units, a segment of
a conical mirror and samples of the radiator materials was built and its
performance was evaluated. Results from the last test
beam performed with ion fragments resulting from the collision of a 158\,GeV/c/nucleon primary beam of indium ions (CERN SPS) on a lead target are
reported. The large amount of collected data allowed to test and
characterize different aerogel samples and the sodium fluoride radiator. In
addition, the reflectivity of the mirror was evaluated. The data analysis confirms the design goals.
\vspace{-0.5cm}
\end{abstract}

\maketitle
\vspace{-1.0cm}
\section{The AMS02 detector}
AMS~\cite{bib:ams} (Alpha Magnetic Spectrometer) is a precision spectrometer 
designed to search for cosmic antimatter, dark matter and to study the relative abundance of elements and isotopic composition of the primary cosmic rays. 
It will be installed in the International Space Station (ISS), where
it will operate at least for three years.

\begin{figure}[htb]
\begin{center}
\vspace{0.2cm}
\epsfig{file=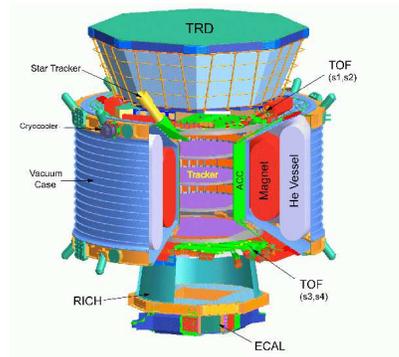,width=0.71\linewidth,clip=,bbllx=0,bblly=0,bburx=612,bbury=550}  
\vspace{-0.5cm}
\caption{A whole view of the AMS Spectrometer.}
\end{center}  
\label{fig:ams}
\vspace{-1.0cm}
\end{figure}

The spectrometer  will be capable of measuring the rigidity ($R\equiv pc/ |Z| e$), the charge ($Z$),
the velocity ($\beta$) and the energy ($E$) of cosmic rays within a 
geometrical acceptance of $ \sim$0.5\,m$^2$sr.
Fig. \ref{fig:ams} shows a schematic view of the AMS spectrometer. The system
is composed of several subdectors: Transition Radiation Detector (TRD),
Time-of-Flight (TOF), Silicon Tracker (STD), Anticoincidence Counters (ACC),
superconducting magnet, Ring Imaging \CK\ detector and electromagnetic
calorimeter (ECAL).

\section{The AMS RICH detector}
The RICH is a proximity focusing device with a dual radiator configuration on
the top made of 92 aerogel 25\,mm thick tiles with a refractive index 1.050 and 
sodium fluoride (NaF) tiles with a  thickness of 5\,mm in the center
covering an area of 34$\times$34\,cm$^2$.  
The NaF placement prevents the loss of photons in the hole existing in the center of the readout plane  
($64\times64$ cm$^2$), in front of the ECAL calorimeter located below.

The detection matrix is composed of 680 multipixelized photon readout cells
each consisting of a photomultiplier coupled to a light guide, HV divider plus
front-end (FE) electronics, all housed and potted in a plastic shell and then
enclosed in a magnetic shielding. The photon detection is made with an array
of multianode Hamamatsu tubes (R7600-00-M16) coupled to a light guide. The
effective pixel size is 8.5\,mm.  

\begin{figure}[htb]
\begin{center}
\vspace{-.5cm}
\begin{tabular}{cc}  
\hspace{-.68cm}
\includegraphics[scale=0.21,angle=0,clip=,bb=14 -24 599 396]{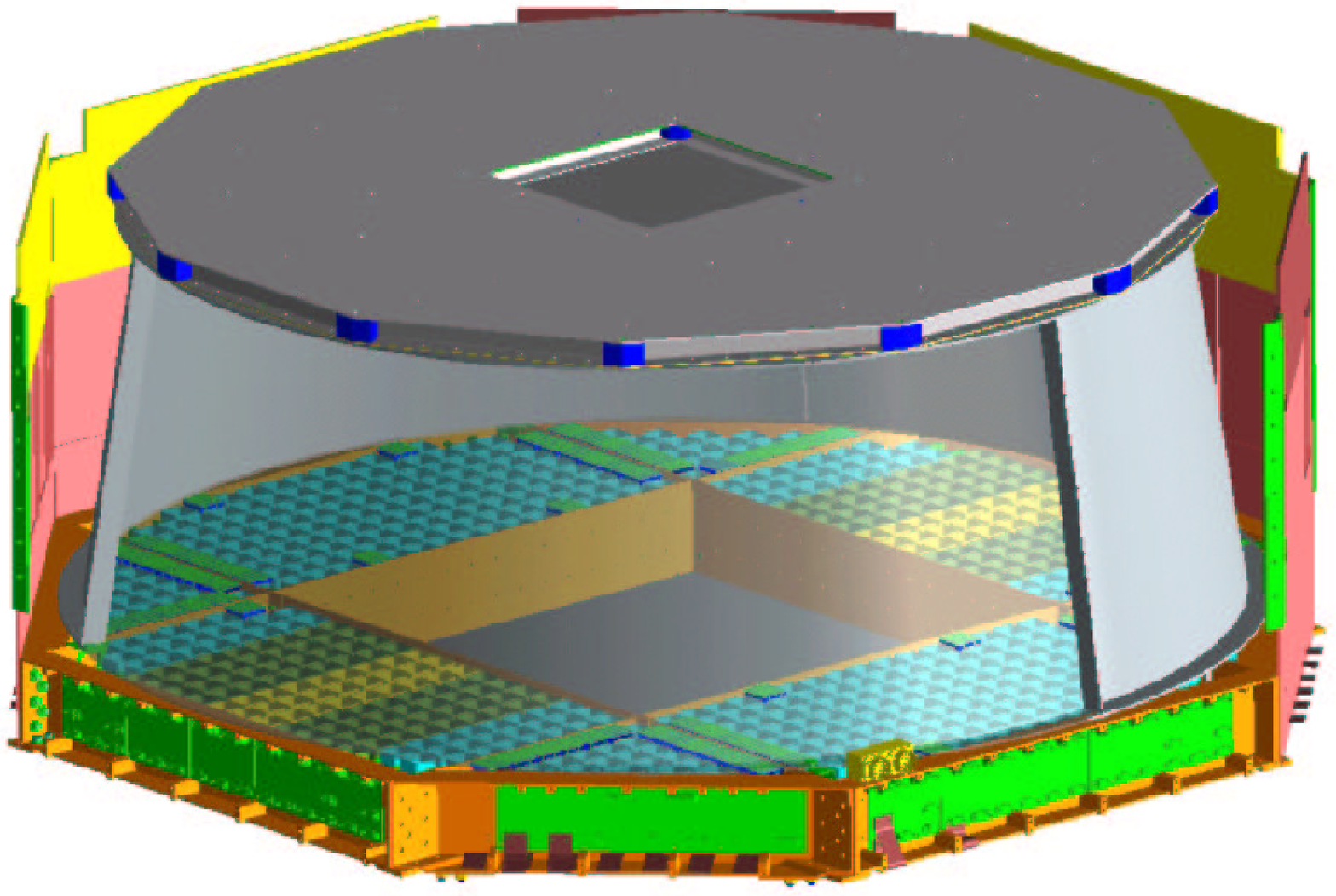}
&
\hspace{-.6cm}
\includegraphics[scale=0.23,angle=0,clip=,bb=14 14 496 463]{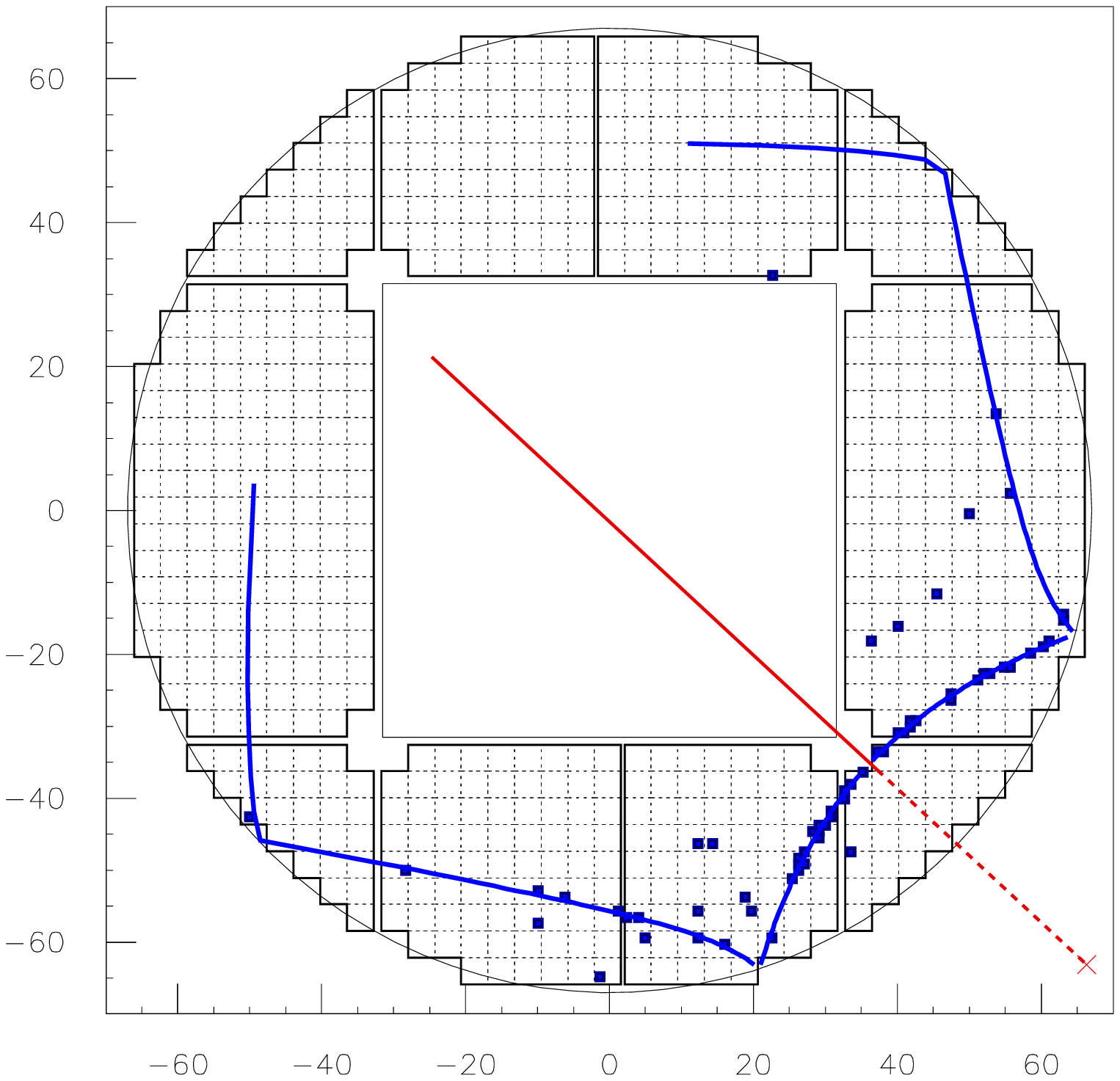} 
\end{tabular}
\vspace{-1.0cm}
\caption{\emph{On the left:} View of the RICH detector. \emph{On the
  right:} Beryllium event display generated in a NaF radiator. \label{fig:rich}}
\end{center} 
\end{figure}

A high reflectivity conical mirror surrounds the whole set. It consists of a
carbon fiber reinforced composite substrate with a multilayer coating made of aluminium
and SiO$_2$ deposited on the inner surface. This ensures a reflectivity
higher than 85\% for 420\,nm wavelength photons.
\mbox{Figure \ref{fig:rich}} (left) shows a schematic view of the RICH detector. 

RICH was designed to measure the velocity
($\beta\equiv v/c$) of singly charged particles with a resolution $\Delta\beta/\beta$ of 0.1$\%$,
to extend the charge separation (Z) up to iron (Z=26), to contribute to $e/p$ separation 
and to albedo rejection.

\begin{figure}
\begin{center}

\begin{tabular}{cc}
\hspace{-0.2cm}
\scalebox{0.21}{%
\includegraphics{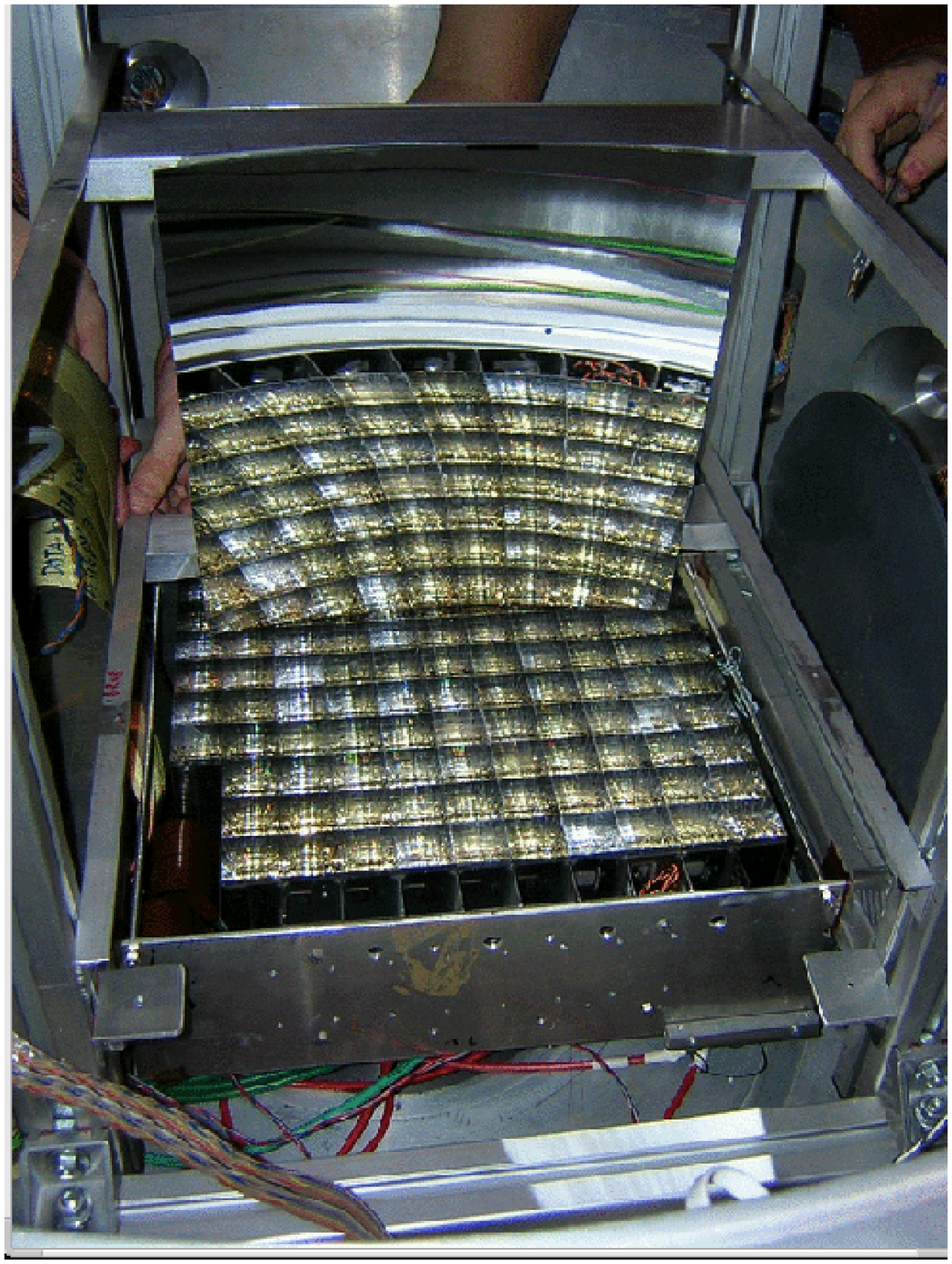}}
&
\hspace{-0.4cm}
\scalebox{0.22}{%
\includegraphics{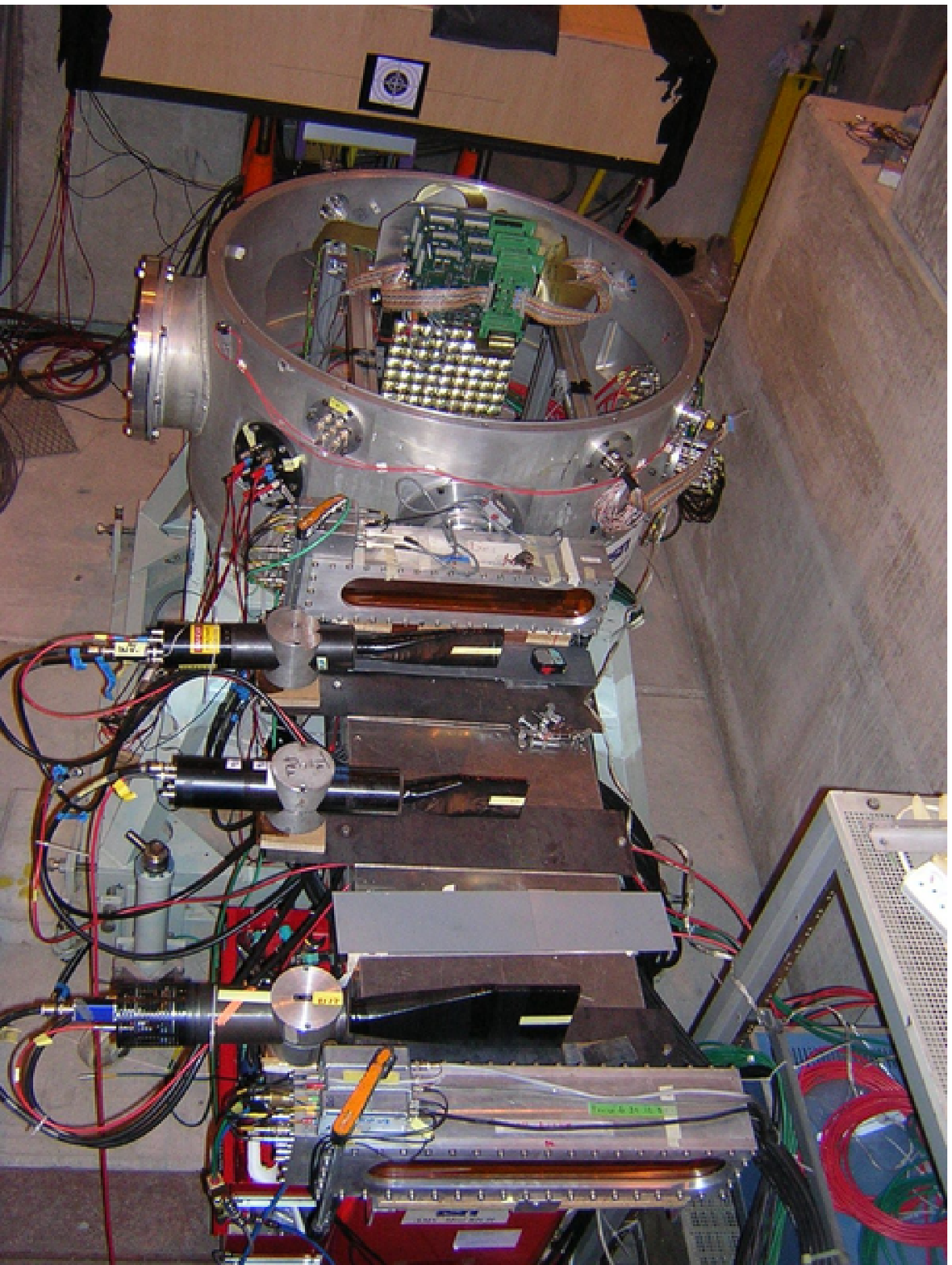}}
\vspace{-0.5cm}
\end{tabular} 
\end{center}
\vspace{-0.5cm} 
\caption{Protype with reflector (left). Top view of the test beam 2003
  experimental setup using CERN SPS facility (right).\label{fig:tb03}}
\vspace{-0.3cm} 
\end{figure}

In order to validate the RICH design, a prototype with an array of
9$\times$11 cells filled with 96 photomultiplier readout units similar to part of the matrix of the final model was constructed. The
performance of this prototype has been tested with cosmic muons and with a
beam of secondary ions at the CERN SPS produced by fragmentation of a primary
beam in 2002 and 2003. The light guides used were prototypes with a slightly
smaller collecting area (31$\times$31\,mm$^2$).
 Different samples of the radiator materials were tested and placed at an
 adjustable supporting structure. Different expansion heights were set
 in order to have fully contained photon rings on the detection matrix like
 in the flight design.
A segment of a conical mirror with 1/12 of the final azimuthal coverage, which is shown in left picture of \mbox{Figure
  \ref{fig:tb03}}, was also tested. 

The RICH assembly has already started at CIEMAT in Spain and is foreseen to
be finished in July 2007. A rectangular grid has already been assembled and has been
subject to a mechanical fit test, functional tests, vibration tests and
vacuum tests. The other grids will follow. The refractive index of the
aerogel tiles is being measured and the radiator container was subjected to a
mechanical test. The final integration of RICH in AMS will take place at CERN
in 2008.

\vspace{-0.8cm} 
\section{Velocity ($\beta$) and charge ($Z$) reconstruction}
\label{BETAZ}
A charged particle crossing a dielectric material of refractive index $n$,
with a velocity $\beta$, greater than the speed of light in that 
medium emits photons.
The aperture angle of the emitted photons with respect to the 
radiating particle track is known as the \CK\ angle, $\theta_c$,
and it is given by $\cos\theta_c = \frac{1}{\beta~n}$ (see \cite{bib:rich}).

It follows that the velocity of the particle, $\beta$, is straightforward
derived from the \CK\ angle reconstruction, which is based on a fit to the
pattern of the detected photons.
Complex photon patterns can occur at the detector plane 
due to mirror reflected photons, as can be seen on right display of Figure
\ref{fig:rich}. The event displayed is generated by a simulated beryllium
nuclei in a NaF radiator. 

The \CK\ angle reconstruction procedure 
relies on the information of the particle direction 
provided by the tracker.
The best value of $\theta_c$ will result from the maximization of a likelihood function, built as the product of the probabilities, $p_i$, that the detected hits belong to a given (hypothesis)
\CK\ photon pattern ring,
\begin{equation}
L(\theta_c) = \prod_{i=1}^{nhits} p_i^{n_{i}}  \left[ r_i(\theta_c) \right].
\label{eq:likelihood}
\end{equation}

Here $r_i$ is the closest distance of the hit to the \CK\ pattern and $n_i$
is the hit signal. For a more complete description of the method see \cite{bib:NIM}.

The \CK\ photons produced in the radiator are uniformly emitted along the
particle path inside the dielectric medium, $L$, and their number per unit of energy ($N$)
depends on the particle's charge, $Z$, and velocity, $\beta$, and on 
the refractive index, $n$. Therefore electric charge ($Z$) is determined from the signal evaluation and
taking into account the different detection efficiencies.

\begin{center}
\begin{equation}
N \propto Z^2 ~\Delta L ~\left( 1-\frac{1}{\beta^2 n^2} \right)
\end{equation}
\end{center}

\subsection{Results with the RICH prototype}
The large amount of collected data in the last test beam at CERN, in October
2003, performed with ion fragments resulting from the collision of a 158
GeV/c/nucleon primary beam of indium ions (CERN SPS) on a lead target,
allowed to test the beta and charge reconstruction algorithms, as well as to
characterize the used radiators \cite{bib:tbnim}. Figure
\ref{fig:tb03} (right) shows a general view of the 2003 test beam setup in the experimental area H8-SPS at CERN.  

\begin{figure}
\begin{center}
\vspace{-0.5cm}

\begin{tabular}{cc}
\hspace{-0.5cm}
\scalebox{0.26}{%
\begin{overpic}[bb=0 0 482 482]{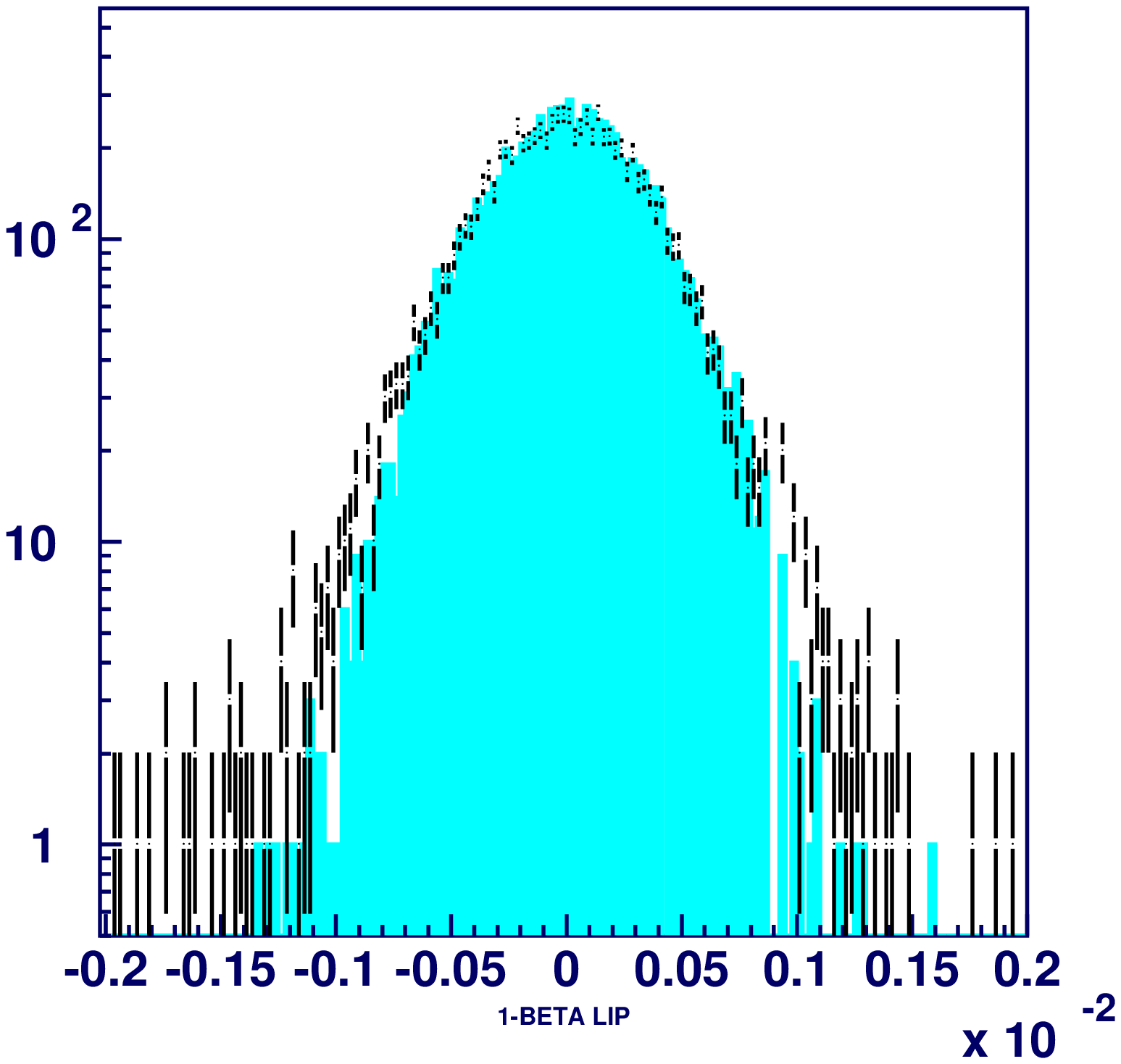} 
\put(67,75){\huge{$\bullet$}}
\put(72,75){\huge{DATA}}
\put(65,70){\myframebox{.03}{cyan}{~}}
\put(72.,70){\huge{MC}}
\end{overpic}
}
&
\hspace{-1.3cm}
\scalebox{0.26}{\includegraphics[bb=0 0 482 482]{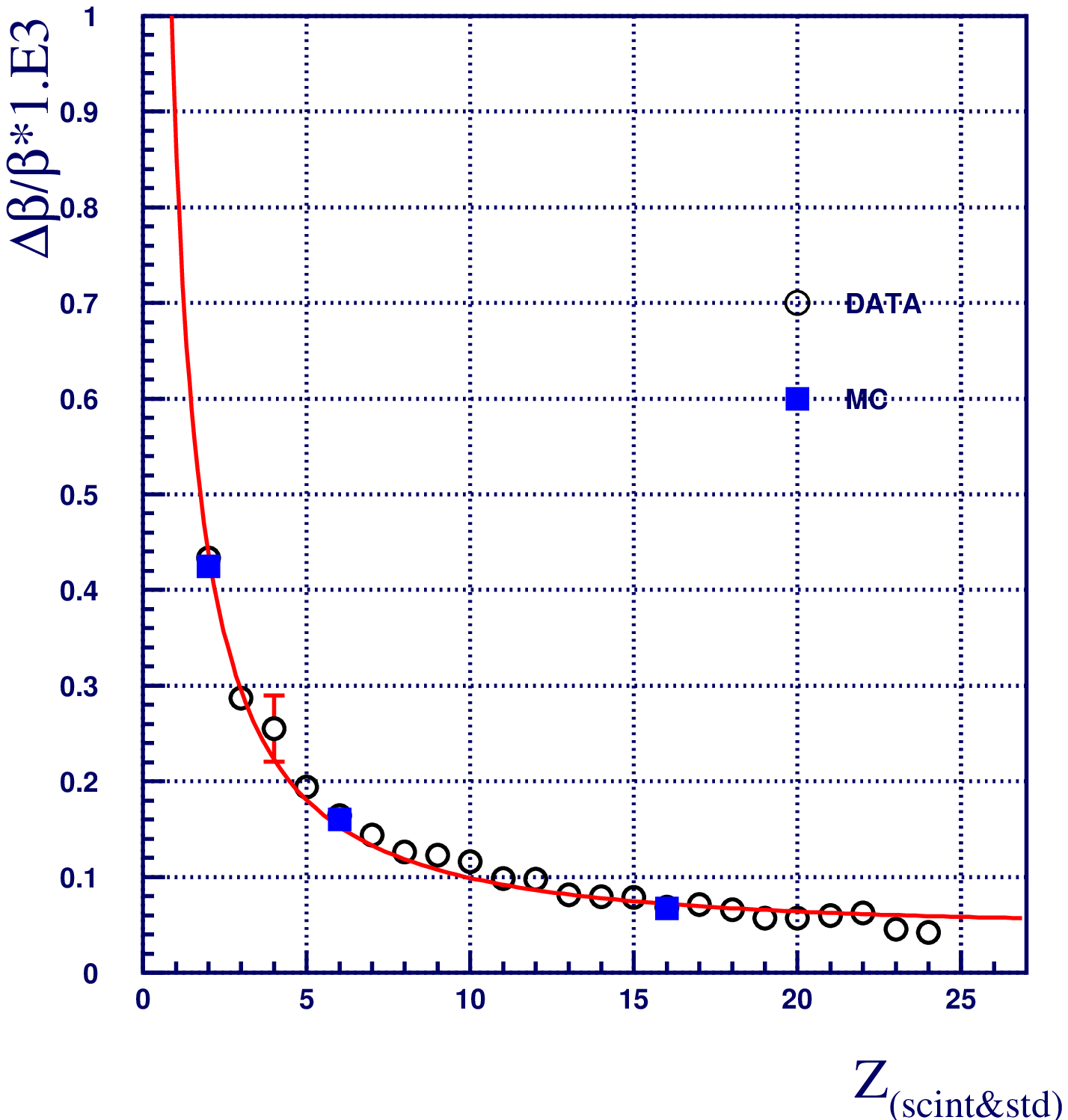}}
\end{tabular}  
\vspace{-0.8cm}
\caption{Comparison of the $(\beta-1)*10^3$ distribution for helium data
  (black dots) and simulation (shaded) (left). Evolution of the $\beta$ resolution with the
  charge obtained for the same aerogel radiator. Simulated points for Z=2, 6,
  16 are marked with full squares (right). \label{fig:betatb03}}
\vspace{-0.5cm}
\end{center}
\end{figure}

The resolution of the $\beta$ measurement, obtained as explained in
\mbox{Section \ref{BETAZ}}, was estimated using a Gaussian fit
to the reconstructed $\beta$ spectrum, shown in left plot of \mbox{Figure
  \ref{fig:betatb03}} for helium nuclei. Data were collected with the aerogel
radiator n=1.05, 2.5\,cm thick together with an expansion height of
35.31\,cm. The events shown correspond to particles impinging vertically and
generating fully contained rings. The beta reconstructed from simulated
helium data is also shown superimposed with a good agreement between data and Monte Carlo (MC). 

The charge dependence of the velocity relative resolution for the same
radiator is shown in the right plot of \mbox{Figure \ref{fig:betatb03}}. The observed resolution varies
according to a law $\propto$1/Z, as it is expected from the charge dependence
of the photon yield in the \CK\ emission, up to a saturation limit set by the
pixel size of the detection unit cell. The function used to perform the fit
is the following:
\begin{equation}
\sigma\left(\beta\right)=\sqrt{\left(\frac{A}{Z}\right)^2+B^2}
\end{equation}
where, $A$ means the $\beta$ resolution for a singly charged particle while $B$
means the resolution for a very high charge generating a large number
of hits.
The fitted values are $A=0.872\pm0.003$ and $B=0.047\pm0.001$ for the run
conditions stated above. 
Simulated data points for Z=2, 6, 16 are marked upon the same plot with full
squares. Once more the agreement between data and MC measurements
for charges different of Z=2 is good.

\begin{figure}
\begin{center}
\vspace{-0.5cm}
\begin{tabular}{cc}
\hspace{-0.3cm}
\scalebox{0.21}{%
\includegraphics[bb=0 0 485 507]{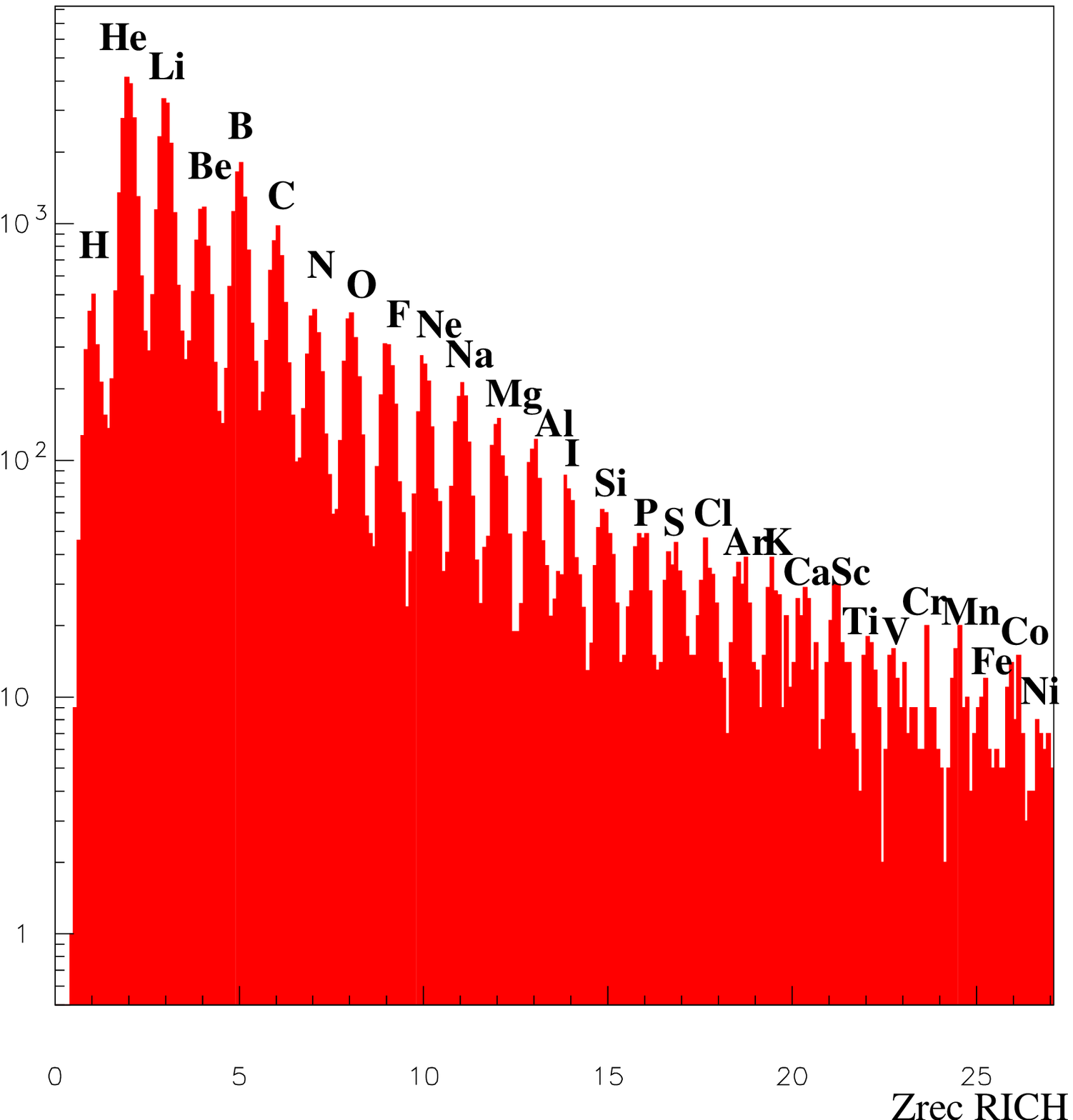} 
}
&
\hspace{-0.5cm}
\scalebox{0.21}{\includegraphics[bb=20 0 567 567]{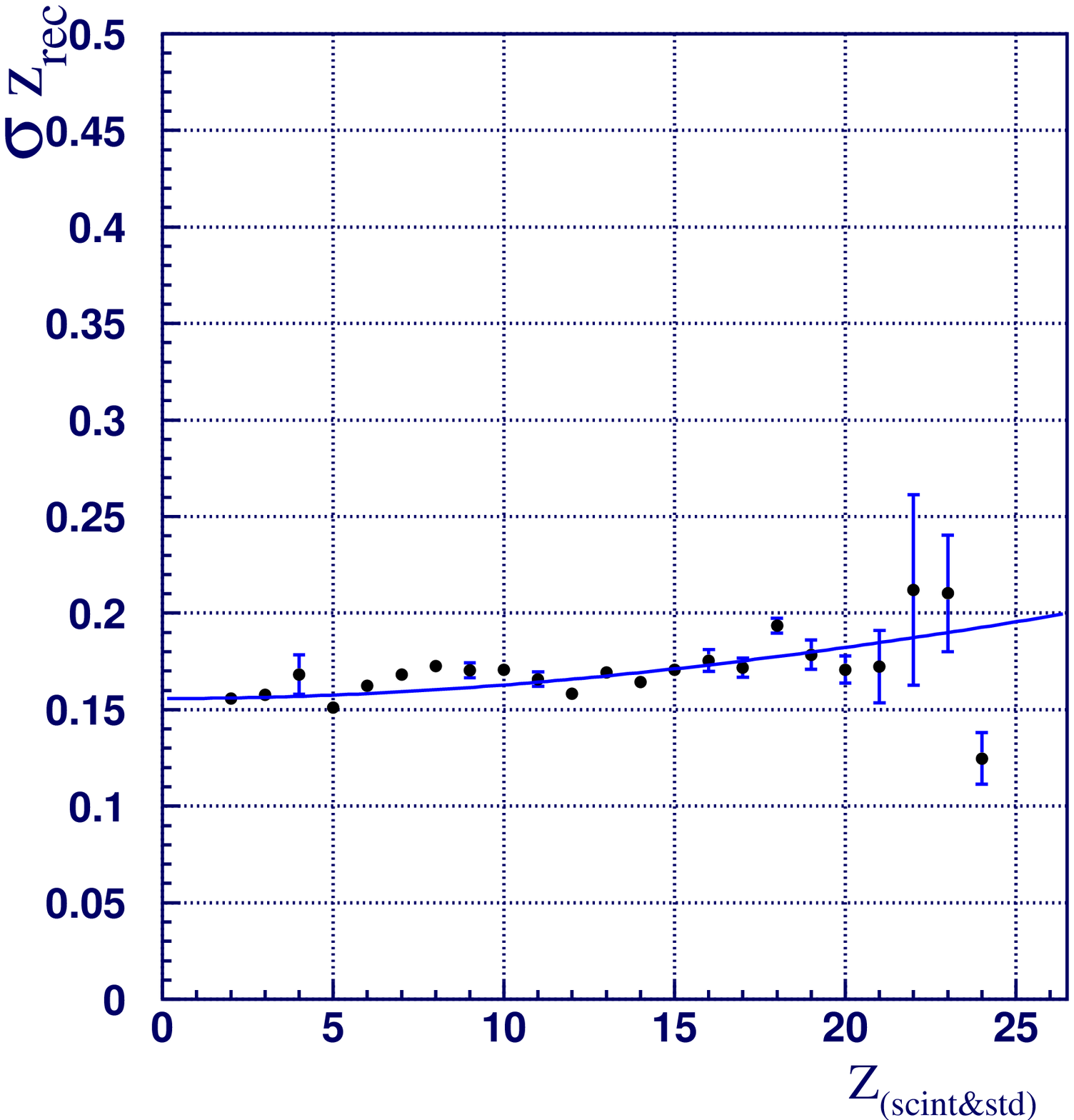}}
\end{tabular}  
\vspace{-0.4cm}
\caption{Charge peaks distribution measured with the  RICH prototype using
  a n=1.05 aerogel radiator, 2.5\,cm thick. Individual peaks are identified
  up to Z$\sim$28 (left). Charge resolution versus particle Z for the same
  aerogel radiator. The curve gives the expected value estimated as explained
  in the text (right). \label{fig:chgtb03}}
\vspace{-0.9cm}
\end{center}
\end{figure}

The distribution of the reconstructed charges in an aerogel radiator of n=1.05,
2.5\,cm thick is shown in left plot of \mbox{Figure \ref{fig:chgtb03}}. The spectrum enhances a structure of well separated
individual charge peaks over the whole range up to Z=28.

The charge resolution for each nuclei, shown in right panel of \mbox{Figure
\ref{fig:chgtb03}}, was evaluated through individual
Gaussian fits to the reconstructed charge peaks selected by the independent
measurements performed by two scintillators and silicon tracker detectors.
A charge resolution for proton events slightly better than $0.17$ charge
units is achieved.

The charge resolution as function of the charge Z of the particle follows a
curve that corresponds to the error propagation on Z which can be expressed
as:
\begin{equation}
\sigma(Z)=\frac{1}{2}\sqrt{\frac{1+\sigma_{pe}^2}{N_0}+Z^2\left(\frac{\Delta N}{N}\right)_{syst}^2}.
\end{equation}
This expression describes the two distinct types of uncertainties that affect Z
measurement: the statistical and the systematic. The statistical term is independent of the
nuclei charge and depends essentially on the amount of \CK\ signal
detected for singly charged particles ($N_{0} \sim 14.7$) and on the
resolution of the single photoelectron peak ($\sigma_{pe}$). The systematic
uncertainty scales with Z, dominates for higher charges and is around 1\%. It
appears due to non-uniformities at the radiator level or at the photon
detection. In order to keep the systematic uncertainties
below 1\%, the aerogel tile thickness, the refractive index and the
clarity should not have a spread greater than 0.25\,mm, $10^{-4}$ and $5\%$, respectively; at the detection level a precise knowledge ($<\,5\%$ level) of the single unit cell photo-detection efficiency and gains is required. 

Runs with a mirror prototype were also performed and its reflectivity was
derived from data analysis. The obtained value is in good agreement with the
manufacturer value.

\section{Conclusions}
AMS-02 will be equipped with a proximity focusing RICH enabling velocity measurements with a resolution of about 0.1\% and extending the charge measurements up to the iron element.
Velocity reconstruction is made with a likelihood method. Charge reconstruction is made in an event-by-event basis. 
Evaluation of both algorithms on real data taken with in-beam tests
at CERN, in October 2003 was done.
The detector design was validated and a refractive index 1.05 aerogel was
chosen for the radiator, fulfilling both the demand for a large light yield
and a good velocity resolution. The RICH detetector is being constructed and
its assembling to the AMS complete setup is foreseen for 2008.

\end{document}